\documentstyle[aps,amsfonts,prbbib]{revtex}

\newfont{\bfsym}{\ifpreprintsty cmbsy10 scaled 1200 \else cmbsy10\fi}

\draft

\def\bftimes{\mbox{\bfsym\char2}}

\newcommand{\ii}{\mbox{\rm i}} 
 
\newcommand{\req}[1]{(\ref{#1})}
\newcommand{\sref}[1]{Sec.~\ref{#1}}
\newcommand{\ie}{{\em i.e.}} \newcommand{\eg}{{\em e.g.}}

\newcommand{\wek}[2]{\newcommand{#1}{\mbox{\bf #2}}}
\wek{\vA}{A} \wek{\vR}{R} \wek{\vH}{H}
\wek{\vp}{p} \wek{\rr}{r} \wek{\va}{a}
\wek{\vt}{t} \wek{\vv}{v} \wek{\vm}{m}
\wek{\vy}{y} \wek{\vx}{x} \wek{\vz}{z}

\begin{document}

 \title{The role of a form of vector potential --- normalization of the
antisymmetric gauge} 
 \author{Wojciech FLOREK~\cite{WSF} and Stanis{\l}aw
WA{\L}CERZ~\cite{StW}} 
 \address{A.~Mickiewicz University, Institute of Physics,
Computational Physics Division, ul.~Umultowska~85, 61--614 Pozna\'n,
Poland}
 \date{July 30, 1997}
\preprint{To be published in J. Math. Phys.}

\maketitle

 \begin{abstract}
 Results obtained for the antisymmetric gauge $\vA=[Hy,-Hx]/2$ by
Brown and Zak are compared with those based on pure group-theoretical
considerations and corresponding to the Landau gauge $\vA=[0,Hx]$.
Imposing the periodic boundary conditions one has to be very careful
since the first gauge leads to a factor system which is not
normalized. A period $N$ introduced in Brown's and Zak's papers
should be considered as a {\em magnetic}\/ one, whereas the {\em
crystal}\/ period is in fact $2N$. The `normalization' procedure
proposed here shows the equivalence of Brown's, Zak's, and other
approaches. It also indicates the importance of the concept of {\em
magnetic cells}. Moreover, it is shown that factor systems (of
projective representations and central extensions) are
gauge-dependent, whereas a commutator of two magnetic translations is
gauge-independent. This result indicates that a {\em form}\/ of the
vector potential (a gauge) is also important in physical
investigations.
 \end{abstract} 

\pacs{PACS numbers: 02.20-a, 03.65.Bz}

\section{Introduction}\label{Intr}

The discovery of the quantum Hall effect~\cite{qhe,huck} led to
remarkable interest in two-dimensional electron systems subjected to
a magnetic field.~\cite{2DES} Since 1980 authors working in different
fields --- from applied to mathematical physics --- have considered
related problems and many new features have been observed and
discussed.~\cite{many} One of the most interesting questionsis the
dynamic of two-dimensional electrons in a periodic potential and an
external magnetic field.~\cite{period} The first results, in the
tight binding approximation, were presented by Peierls,~\cite{Pei}
shortly after Landau's~\cite{Lan} discovery of the quantization of
electron states in a magnetic field. A new impact was due to
Brown~\cite{brown} and Zak~\cite{zak1a,zak1b} who independently
introduced magnetic translation operators in two different, but
equivalent, ways. Both approaches were based on group-theoretical
considerations and led to the broadening of the Landau levels and
quantization of a magnetic field.~\cite{brown,zak1c} Although more
than thirty years have passed, their papers are still considered as
fundamental ones.~\cite{period} Brown and Zak proved that the
problem considered is in fact two-dimensional and their
investigations confirmed the importance of projective representations
and central extensions in quantum physics.~\cite{Divx2} On the other
hand, Zak's and Brown's results were not gauge-independent --- only a
completely antisymmetric vector potential was considered by both
authors. An attempt to consider gauge-equivalent vector potentials
leads to some ambiguities and misconceptions if it is not done
carefully. A bit simpler and more clear results can be obtained from
pure group-theoretical considerations. For example, Divakaran and 
Rajagopal did not consider gauges at all and they worked with central
extensions and projective representations only.~\cite{DivRaj}
However, pure mathematical description may not provide us with an
intuitive image of the physical phenomena. Moreover, many experiments
and theories indicate the importance of vector potential,~\cite{AB}
so it is necessary to include gauges and potentials in
considerations. 

The aim of this paper is to show sources of misconceptions,
ambiguities, and unexpected gauge-dependence of the problem. In
particular, factor systems of projective representations and central
extensions introduced by Brown and Zak have been carefully checked
and compared with those obtained from pure group-theoretical
considerations.~\cite{DivRaj,florek94,florek96a} It occurs that they
can be considered as standard but they are not
normalized.~\cite{Divx2,alt} This last fact is the main source of
differences between Brown's and Zak's approaches. Moreover, it
indicates points at which a form of the vector potential is
important, \ie, the points at which the problem is not
gauge-independent.

In this paper we propose a procedure of `normalization' of those
factor systems, which enables us to identify irreps introduced by
Brown and Zak. A comparison of these irreps with those obtained for
central extensions of finite translation groups leads to a concept of
the so-called {\em magnetic cells}~\cite{zak1b} and shows that Brown
and Zak considered in fact finite lattices with a period $2N$ not $N$.

For the sake of clarity, the following simplifications arising from
the quoted papers are assumed. Position (\rr, \vR), momentum (\vp),
and vector potential (\vA) are considered to be two-dimensional
vectors. Note that $\rr=(x,y)$ is any vector of ${\Bbb R}^2$, whereas
$\vR=(X,Y)\in{\Bbb Z}^2$ denotes a vector of a square lattice with
$\va_1=\hat{\vx}$ and $\va_2=\hat{\vy}$, so the area of the
elementary cell is equal to 1. The magnetic field is perpendicular to
the $x$-$y$ plane and $\vH=H\hat{\vz}$. The periodic boundary
conditions are imposed on representations of ${\Bbb Z}^2$ and the
periods are equal, \ie, $N_1=N_2=N$; the finite translation group and
its representations can be considered equivalently.

The paper is organized as follows. In \sref{Des} the most fundamental
formulas of Brown's and Zak's papers are recalled and equivalence of
their approaches are indicated. Basic properties of projective
representations are briefly presented, too. The role of factor
systems is briefly discussed in \sref{Stan}. The next section is
devoted to determination of the equivalence of different approaches.
From the physical point of view it is done by introducing the concept
of magnetic cells. The results obtained are discussed in \sref{Dis}.

\section{Different descriptions of magnetic translation
groups}\label{Des}

From the algebraic point of view there are two equivalent
descriptions of the magnetic translation operators.
Brown~\cite{brown} investigated a {\em projective}\/ representation
of the translation group $T$ then imposed the magnetically periodic
boundary conditions on it. On the other hand, Zak~\cite{zak1a}
introduced a closed set of noncommuting operators which, in fact,
form a covering group $T'$ of $T$ so its standard (vector)
representations are {\em projective}\/ representations of
$T$.~\cite{alt} The finiteness of these representations was again
achieved by imposing the periodic boundary conditions. These two
approaches are related by a formula which follows from the induction
procedure if one constructs representations of the covering group.
Since $T'$ is a central extension of $T$ by the group $U(1)$ (or its
subgroup referred to hereafter as a group of factors and denoted
$F$)~\cite{florek94,florek96a} then its (vector) representations can
be written as
 \begin{equation}\label{xidef}
 \Xi[\alpha,\vR]=\Gamma(\alpha)D(\vR)\,,
 \end{equation}
 where $\alpha\in F\subset U(1)$, $\vR\in T$, $\Gamma$ is a vector
representation of $F$ and $D$ is a {\em projective}\/ representation
of $T$. A factor system $m(\vR,\vR')$ of this representations is
determined by the relation~\cite{alt}
 \begin{equation}\label{facdef}
 D(\vR)D(\vR')=m(\vR,\vR')D(\vR+\vR')\,,
 \end{equation}
 whereas the multiplication rule for $T'$ reads
 \begin{equation}\label{mulrul}
 [\alpha,\vR][\alpha',\vR']=[\alpha\alpha'\mu(\vR,\vR'),\vR+\vR']
 \end{equation}
 with $\mu(\vR,\vR')$ being a factor system of a central extension.
These factor systems are related to each other by the formula 
 \begin{equation}\label{facrel}
 m(\vR,\vR')=\Gamma[\mu(\vR,\vR')]\,.
 \end{equation}
 This relation establishes the equivalence of both approaches.
Moreover, both authors assumed the antisymmetric vector potential
(gauge) $\vA=(\vH\bftimes\rr)/2=[Hy,-Hx]/2$ and were not able to
generalize their considerations to other gauges, in particular their
approaches did not include the Landau gauge. On the other hand, their
results and some conclusions are different in some points which will
be discussed here~\cite{sign} and compared with the results obtained
for the Landau gauge. 

All considerations and formulas given above are also valid for a
finite group $T_N$ and its (finite-dimensional) representations. In
addition, we can apply to this case a version of the Burnside
theorem which reads that nonequivalent irreducible projective
representations of $T_N$ with the same factor system $m(\vR,\vR')$
satisfy the following condition~\cite{alt}
 \begin{equation}\label{Burn}
 \sum_j [\,{^j}\!D]^2=|T_N|=N^2\,,
 \end{equation}
 where $j$ labels nonequivalent representations (there is no
expression for a number of these representations) and $[\,{^j}\!D]$
denotes the dimension of $\,{^j}\!D$. Since $F$ is an Abelian group
then it has $|F|$ irreducible nonequivalent representations and each
of them determines different (nonequivalent) factor system
$m(\vR,\vR')$ according to \req{facrel}. It follows from \req{Burn}
that irreducible representations of $T'_N$ determined by \req{xidef}
satisfy the Burnside theorem.

Brown~\cite{brown} defined a magnetic translation operator as
 \begin{equation}\label{Bmto}
 \widehat{T}(\vR)=\exp[-\ii\vR\cdot(\vp-e\vA/c)/\hbar]\,,
 \end{equation}
 where $\vp$ is the kinetic momentum and $\vA$ is the vector
potential such that $\nabla\bftimes\vA=\vH$. These operators form a
projective representation of $T$ with a factor system~\cite{brown} 
 \begin{equation}\label{Bfac}
 m(\vR,\vR')=\exp[-\pi\ii(\vR\bftimes\vR')\cdot \vH/\varphi_0]
 \end{equation}
 where $\varphi_0=ch/e$. Brown showed that one can impose the
periodic boundary conditions $\widehat{T}(N\va_j)\psi=\psi$ if
(notice simplifications assumed in this paper --- in fact, $H$
denotes hereafter the magnetic flux through one primitive cell)
 \begin{equation}\label{Bcon}
 H = {l\over N}\varphi_0\,,
 \end{equation}
 where $l$ is mutually prime with $N$, \ie\ $\gcd(l,N)=1$. Hence a
factor system of a finite projective representation $\,{^l}\!D$ for
$H$ satisfying \req{Bcon} is given as
 \begin{equation}\label{BZfac}
 m_l([X,Y],[X',Y']) = \exp[-\pi\ii l (XY'-YX')/N]\,.
 \end{equation}
 Brown showed that there is the unique (up to equivalence)
irreducible projective representation with a dimension $N$ and matrix
elements~\cite{brown}
 \begin{equation}\label{Brep}
 {^{\,l}_1}D_{jk}[X,Y]=\exp
 \biggl[\pi\ii\frac{lX}{N}(Y+2j)\biggr]\delta_{j,k-Y}\,,
 \end{equation}
 where $j,k=0,1,\dots,N-1$ and $\delta_{j,k}$ is calculated modulo
$N$ (Brown labeled rows and columns by $j,k=1,2,\dots,N$).

Zak~\cite{zak1a} considered a covering group of the translation group
consisting of operators
 \begin{equation}\label{Zmto}
 \tau(\vR|\vR_1,\dots,\vR_j)=\widehat{T}(\vR)
 \exp[2\pi\ii\phi(\vR_1,\dots,\vR_n)/\varphi_0]\,,
 \end{equation}
 where $\sum_i\vR_i=\vR$ and $\phi(\vR_1,\vR_2,\dots,\vR_j)$ is the
flux of the magnetic field through a polygon enclosed by a loop
consisting of the vectors $\vR_1,\vR_2,\dots,\vR_j,-\vR\,$. The
periodicity condition was the same as \req{Bcon} for even $N$ but for
odd $N$ Zak proved that the condition
 \begin{equation}\label{Zcon}
 H = 2{l\over N}\varphi_0
 \end{equation}
 should be satisfied. This condition implies that for even $N$
the number of different factors is $2N$, whereas it equals $N$ for
odd $N$.~\cite{zak1a,zak1b} The result obtained by Zak agreed with
Azbel's considerations~\cite{azbel} who showed that wave functions
had to be periodic functions of $H$ with the period $2\varphi_0$. It
is worthwhile noting that Azbel also worked with the antisymmetric
gauge. Zak did not introduce a factor system in an explicit way (it
was not necessary in his constructions of representations) but it can
be easily found by considering multiplication of coset
representatives $\tau(\vR|\vR)$,~\cite{alt,BR} which simply are equal
to $\widehat{T}(\vR)$ [see \req{Zmto}]. Therefore the factor system
is also given by \req{BZfac}, but now it is the factor system of the
covering group being a central extension so it should be denoted as
$\mu_l(\vR,\vR')$. Zak~\cite{zak1b} also showed that matrix elements
of an irreducible $N$-dimensional representation should be (only the
coset representatives $\tau(\vR,\vR)$ are taken into account
here)~\cite{m3n3}
 \begin{equation}\label{Zrep}
{_2^{\,l}}D_{jk}\Bigl[\tau([X,Y]|[X,Y])\Bigr]=\exp
 \biggl[2\pi\ii\frac{lX}{bN}(Y+2k)\biggr]\delta_{j,k+Y}\,,
 \end{equation}
 where $b=1,2$ for $N$ odd and even, respectively. It is obvious that
this representation corresponds to the irreducible representation
$\Gamma(\alpha)=\alpha$ of the factor group $F$, so
$m_l(\vR,\vR')=\mu_l(\vR,\vR')$. According to \req{Bcon} and
\req{Zcon} changes of $H$ are related to changes of $l$ but they were
interpreted in different ways. In Brown's considerations $H$
determines a factor system of projective representations in a direct
way --- different values of $H$ satisfying \req{Bcon} lead to
nonequivalent projective representations. On the contrary, Zak
considered different (nonequivalent) central extensions of $T$ with
factor systems $\mu_l$. However, Zak assumed that only the
representations $\Gamma(\alpha)=\alpha$ were physical whereas the
others were rejected as nonphysical in further
considerations.~\cite{zak1b,florek97a} It means, according to Zak,
that for a central extension with a factor system $\mu_l$ one has to
find projective representations $\,{^l}\!D$ with a factor system
$m_l=\Gamma(\mu_l)=\mu_l$. The same result can be obtained while
considering only the factor system $\mu_1$ and next all irreducible
representations $\Gamma_l$ of $F$ such that $\Gamma_l(\mu_1)=m_l$. 
Thus all representations necessary in physical applications,
considered by Zak as representations of different although isomorphic
groups, can be obtained by use of `nonphysical' representations
$\Gamma_l$ with $l>1$. Nevertheless, it seems that the
representations introduced by Brown \req{Brep} could be used in Zak's
approach to construct (vector) representations of $T'$ (finite or
not) according to \req{xidef}. A comparison of \req{Brep} and
\req{Zrep} shows that for odd $N$ Brown and Zak used different
representations. However, for even $N$ ($b=2$) we have
 \begin{equation}\label{BZrep}
 {^{\,l}_1}D_{jk}[X,-Y]=
 {_2^{\,l}}D_{jk}\Bigl[\tau([X,Y]|[X,Y])\Bigr]\,,
 \end{equation}
 where the sign `$-$' originates from a different choice of the sign
of $e$ assumed by Zak~\cite{sign} (in Zak's approach eigenvectors of
$D[1,0]$ are permuted by $D[0,1]$ in the opposite direction than that
assumed in Brown's definition).

The third approach is based on pure group-theoretical considerations
and consists in determination of all possible central extensions of a
finite group ${\Bbb Z}_N^2$ (in general ${\Bbb Z}_{N_1}\otimes{\Bbb
Z}_{N_2}$) by an infinite ($U(1)$) or finite ($C_N=\{\alpha\in {\Bbb
C} \mid\alpha^N=1\}$) group of factors
$F$.~\cite{DivRaj,florek94,florek96a} It was shown, by means of the
Mac~Lane method, that all nonequivalent factor systems corresponding
to finite magnetic translation groups can be written
as~\cite{florek94,florek96a})
 \begin{equation}\label{Ffac}
 \mu_k([X,Y],[X',Y']) = \exp(2\pi\ii kYX'/N)
 \end{equation}
 with $k=0,1,\dots,N-1$. Some important facts have to be mentioned:
 \begin{itemize}
 \item This formula resembles the Landau gauge $\vA=[0,Hx]$; recently
it has been shown that this convergence is not
accidental.~\cite{florek97b}
 \item The fraction $k/N$ can be interpreted as
$H/\varphi_0$,~\cite{florek94,florek96a} so the resulting numbers
constitute a periodic function of $H$ with the period $\varphi_0$ in
agreement with Brown's result but contrary to Zak's and Azbel's 
results.
 \item In both Brown's and Zak's approaches a (group-theoretical)
commutator of two magnetic translations, corresponding to vectors
$[X,Y]$ and $[X',Y']$, is equal to
 \begin{equation}\label{BZcom}
 c([X,Y],[X',Y'])=\exp[-2\pi\ii(XY'-YX')H/\varphi_0]
 \end{equation}
 and it is the same as obtained in the cited
papers~\cite{florek94,florek96a} if $H$ does not satisfy \req{Zcon}
but \req{Bcon}.
 \item There are no additional conditions imposed on $k$ and on $N$
(\ie, the results are valid for both odd and even $N$ and for
$\gcd(k,N)>1$).
 \end{itemize}
 Taking into account only parameters $k=l$ mutually prime with $N$ it
can be shown that $N$-dimensional irreducible projective
representations of $T_N$ (or `physical' vector representations of the
extension of $T_N$ by $C_N$) have the following matrix elements
 \begin{equation}\label{Frep}
 {^{\,l}_3}D_{jk}[X,Y]=\exp
 \biggl(2\pi\ii\frac{l}{N}Xj\biggr)\delta_{j,k-Y}\,.
 \end{equation}
 Representations with $\gcd(k,N)>1$ were briefly discussed
elsewhere,~\cite{florek97a} but the difference between odd and even
$N$ was not considered there.

\section{Projective representations --- standard and normalized
factor systems}\label{Stan}

To compare different descriptions of the magnetic translation groups
we have to discuss not only projective representations themselves but
also their factor systems. To begin with we recall now some
definitions related to factor systems and their
properties.~\cite{alt,BR,CR,kur} As one can see factor systems appear
in the definition of a projective representation \req{facdef} and in
the multiplication rule for a central extension of groups
\req{mulrul}. Factor systems $m_l$ are determined directly (as in
Brown's approach) or via factor systems $\mu_l$ for central
extensions by means of the `physical' representations
$\Gamma(\alpha)=\alpha$ and the formula \req{facrel}.

A factor system $m\colon T\times T\to{\Bbb C}$ has to satisfy the
following condition~\cite{BR,kur}
 \begin{equation}\label{faccon}
 m(\vR,\vR')m(\vR+\vR',\vR'')=m(\vR',\vR'')m(\vR,\vR+\vR'')
 \end{equation}
 for all $\vR,\vR',\vR''$. A {\em trivial}\/ factor system
$t(\vR,\vR')$ is determined by any mapping $f\colon T\to{\Bbb C}$
according to 
 \begin{equation}\label{tfacsys}
 t(\vR,\vR')=f(\vR)f(\vR')/f(\vR+\vR')\,.
 \end{equation}
 Since $T$ is Abelian then each trivial factor system is symmetric,
\ie, $t(\vR,\vR')=t(\vR',\vR)$. If 
 \begin{equation}\label{efacsys}
 m'(\vR,\vR')=t(\vR,\vR')m(\vR,\vR')
 \end{equation}
 then factor systems $m$ and $m'$ are called {\em equivalent}.
Notice, however, that the projective representations determined by
equivalent factor systems are {\em nonequivalent}.~\cite{alt} Since
all factor systems for a given $T$ form an Abelian group $\Phi$ and a
set of trivial factor systems $\Theta$ is its normal subgroup then
elements of the factor group $M=\Phi/\Theta$ (known as the {\em Schur
multiplicator}) correspond to representatives of classes of
equivalent factor systems. A factor system is called {\em standard}\/
if it satisfies
 \begin{equation}\label{sfacsys}
 m(\vR,{\bf0})=m({\bf0},\vR)=1\,,\qquad\forall\, \vR\,.
 \end{equation}
 A factor system of an $N$-dimensional projective representation is
{\em normalized} if 
 \begin{equation}\label{nfacsys}
 m(\vR,\vR')\in C_N\,,\qquad\forall\, \vR\,,\vR'
 \end{equation}
 (\ie, each factor is the $N$th root of 1). 

It is well known that the Schur multiplicator of ${\Bbb Z}_N^2$ is
$C_N$,~\cite{baba} so the factor system \req{Ffac} is normalized and
standard since $m_k([0,0],[X,Y])=1$. Moreover, it is periodic with
respect to $Y$ and $X'$ --- the period is equal to $N$. In particular
we have $m_k([N,0],[X,Y])=m_k([X,Y],[N,N])=1$, {\em etc}. On the other
hand, the factor system \req{BZfac} of $N$-dimensional
representations \req{Brep} or \req{Zrep} is not normalized because
some of factors do not belong to $C_N$ but to $C_{2N}$ instead. It
also means that this system is standard, because
$m_l([0,0],[X,Y])=1$, but it appears that $N$ does not serve as a
period because, for example, $m_l([0,N],[1,0])=\exp(\pi\ii
l)=(-1)^l$. This fact stirs up a conflict between the conditions
obtained by Brown and those obtained by Zak and, moreover, leads to
difficulties in studying magnetic translations for the antisymmetric
gauge. Of course, one may work with factor systems (and, hence,
representations) which are neither standard nor normalized, but such
considerations have to be carried very carefully and results obtained
have to be carefully interpreted, too.~\cite{alt} Brown and Zak did
not check normalization of their factor systems and this led to
ambiguity of their results [{\em cf.}\ \req{Bcon} and \req{Zcon}].

At first let us notice that Brown took into account one requirement
only, namely~\cite{brown}
 \begin{equation}\label{weak} 
\widehat{T}(N\va_j)\widehat{T}(\vR)=\widehat{T}(\vR)\widehat{T}(N\va_j
)\,,
 \end{equation}
 \ie, that $\widehat{T}(N\va_j)$ would commute with any other
operator. On the other hand, Zak demanded in addition
that~\cite{zak1a}
 \begin{equation}\label{strong} 
 m(N\va_j,N\va_k)=1\,,
 \end{equation}
 \ie, that $\,{^l}\!D(N\va_j)$ should behave as a constant factor. As
follows from \req{BZfac} 
 $$
 m_l([N,0],[0,N]) = (-1)^{lN}\,,
 $$
 so for odd $N$ the magnetic field $H$ has to be twice as high as in
\req{Bcon}. Note that representations \req{Frep}, corresponding to the
Landau gauge, satisfy, for both odd and even $N$, the following
stronger condition
 \begin{equation}\label{strongx} 
 {^{\,l}_3}D(N\va_j){^{\,l}_3}D(\vR)={^{\,l}_3}D(\vR+N\va_j)
 ={^{\,l}_3}D(\vR)\,,
 \end{equation}
 \ie, both $N\va_j$ and $\widehat{T}(N\va_j)$ are equal to the unit
element in the translation group $T_N$ and in the group of magnetic
translation operators, respectively. The condition \req{Zcon} (for
odd $N$) removes these problems but, however, leads to another
question why odd and even $N$ should be considered separately while
both cases can be evidently treated as one with the use of the
Landau gauge. 

While considering restrictions imposed on $H$ by the periodic
boundary conditions with the Landau gauge, \ie, the standard factor
system \req{Ffac}, one can see that the condition $H=k\varphi_0/N$ is
sufficient. So, it seems that Brown's approach is well-supported.
Moreover, it should be noted that Zak weakened his requirements and
later on he considered only Brown's condition.~\cite{zak2} To
enlighten the problem we have to check whether the factor systems
\req{BZfac} and \req{Ffac} are equivalent or not. At first note that
the group-theoretical commutator of operators of any projective
representations (of an Abelian group $T$) is equal to 
 \begin{equation}\label{comm}
 c(\vR,\vR')= D(\vR)D(\vR')[D(\vR')D(\vR)]^{-1}
 =\frac{m(\vR,\vR')}{m(\vR',\vR)}\,,
 \end{equation}
 then it is the same for all equivalent factor systems. Since the
equivalence of factor systems means the equivalence of vector
potentials (gauges),~\cite{huck,florek97b} then the above commutator
does not depend on \vA\ but rather on \vH\ and in this sense this
commutator only (not a factor system) has the physical meaning ---
if $D(\vR)$ represents a lattice translation in the presence of a
magnetic field then the commutator corresponds to a loop determined
by vectors $\vR,\vR',-\vR,-\vR'$ and its value depends on the flux
through a nonprimitive, in general, cell determined by these vectors.
So, Brown's requirement \req{weak} leading to the condition
\req{Bcon} was based on a reasonable assumption. However, the
factor system considered was not normalized which led to a
disagreement with Zak's results.
 
\section{Equivalence of factor systems and representations}\label{Eq}

Let us consider a mapping $f_w[X,Y]=\exp(2\pi\ii wXY)$, $w\in{\Bbb
R}$, which determines the following trivial factor system
 \begin{equation}\label{phiw}
 t_w([X,Y],[X',Y']) = \exp[-2\pi\ii w(XY'+YX')]\,.
 \end{equation}
 The factors obtained belong to $C_N$, \ie, $t$ is standard and
normalized, if $w=j/N$. For example for $j=k$ the factor system
\req{Ffac} is transformed to
 \begin{equation}\label{Ffacbar}
 \overline{\mu}_k([X,Y],[X',Y']) =
 (\mu_k\circ t_{k/N})([X,Y],[X',Y']) =
 \exp(-2\pi\ii kXY'/N)\,,
 \end{equation}
 which corresponds to another form of the Landau gauge
$\bar{\vA}=[-Hy,0]$. It is important that if $t$ is not normalized
then a new factor system $m'=tm$ is not normalized, too. This is,
however, the case which leads to the factor system \req{BZfac}
determined by Brown and Zak --- one has to put $w=k/2N$. This, and
the previous discussion on the commutator, proves that the stronger
condition introduced by Zak following from \req{strong} is
superfluous. It can be easily shown for odd $N$, since for $l$
mutually prime with odd $N$ also $\gcd(2l,N)=1$ (the mapping
$l\mapsto2l$ is an automorphism of ${\Bbb Z}_N$ which changes the
order of elements only). Therefore in the formulas obtained by
Brown, (\ref{Bcon})--(\ref{Brep}), one can replace $l<N$ in the
following way
 \begin{equation}\label{repl}
 l=\cases{0, & for $l=0$\cr
 2k=2l', & for even $l\neq0$\cr 
 2k-1=2(k+N')=2l', & for odd $l$,}
 \end{equation}
 where $N'=(N-1)/2$, $k=1,2,\dots,N'$, and $l'=1,2,\dots,2N'=N-1$. In
this way a relation similar to \req{BZrep} is obtained
 \begin{equation}\label{BZrepodd}
 {^{\,l}_1}D_{jk}[X,-Y]=
 {_2^{\,l'}}\!D_{jk}\Bigl[\tau([X,Y]|[X,Y])\Bigr]\,,
 \end{equation}
 where $l$ and $l'$ are interrelated by \req{repl}. In the same way one
can transform the factor system \req{Ffac} into \req{BZfac}. If
$\gcd(l,N)=1$ then $l$ is replaced by $l'$, so
 \begin{equation}\label{Ffacp}
 m_l([X,Y],[X',Y']) = \exp(2\pi\ii (2l')YX'/N)
 \end{equation}
 and next $w$ is taken to be $l'/N$. The factor system obtained
 \begin{equation}\label{Ffacpp}
 t_wm_l([X,Y],[X',Y']) 
 = \exp[\pi\ii l (YX'- XY')/N]
 \end{equation} 
 is exactly the same as \req{BZfac}. In a sense, we have performed a
`normalization' of the factor system used by Brown and Zak. In 
other words, the projective representations \req{Brep} do not satisfy
the condition \req{strong} for odd $N$ since they are not given in
{\em normalized} form. Such a form can be obtained by substitution
$l\to2l'$ which leads to the condition \req{Zcon} determined by Zak.

Anyway, this way of normalization is not possible in the case of even
$N$, since in general $\gcd(l,N)\neq\gcd(2l,N)$. However, we can use
a hint given by Zak, who did not exploit it in full. At the end of
his paper~\cite{zak1a} Zak noticed that a finite magnetic translation
group contains $N^3$ elements~\cite{m3n3} for odd $N$ whereas for
even $N$ the number of elements is two times bigger. It suggests that
a group considered by him was, in fact, an extension of $T_N$ by
$C_{2N}$ --- the factors obtained were not normalized since they did
not belong to the multiplicator of $T_N$. 

In a previous paper~\cite{florek94} it was shown that central
extensions of $T_N$ by $C_{2N}$ which correspond to magnetic
translation groups have factor systems
 \begin{equation}\label{Ffac2}
 _2m_k([X,Y],[X',Y']) = \exp\Bigl(\frac{\pi\ii}{N} 2kYX'\Bigr)
 \end{equation}
 with $k=0,1,\dots,N-1$. A mapping $f_w$ assigns to each $[X,Y]$ an
element of $C_{2N}$ so it is well defined for $w=k/2N$, \ie,
 $$
 f_w[X,Y]=\exp\Bigl(\frac{\pi\ii}{N} kXY\Bigr)\,.
 $$
 Note that the product $kXY$ is calculated modulo $2N$ and, therefore,
$f_w$ is a multivalued function: in ${\Bbb Z}_N$ numbers $X$ and
$X+N$ represent the same element, whereas 
 $$
 f_w[X+N,Y]=(-1)^{kY}\exp\Bigl(\frac{\pi\ii}{N} kXY\Bigr)
 $$
 is not equal to $f_w[X,Y]$, in a general case. To calculate a
trivial factor system $t_w$ according to \req{tfacsys} one has to
determine $f_w(\vR+\vR')$. Let us assume, at this moment only
formally, that a sum of vectors in this formula will {\em not}\/ be
calculated modulo $N$. Then
 \begin{equation}\label{tfac2}
 t_w([X,Y],[X',Y']) =\exp\Bigl[-\frac{\pi\ii}{N} k(XY'+X'Y)\Bigr]
 \end{equation}
 and 
 $$
 _2m_kt_w([X,Y],[X',Y']) = \exp\Bigl[\frac{\pi\ii}{N}
k(YX'-XY')\Bigr]\,,
 $$
 which coincides with \req{BZfac}. It means that in order to obtain
the above results we have to treat $[X,Y]$ as an element of $T_{2N}$
rather than that of $T_N$. So, in fact, we have considered a larger,
$2N\times2N$, lattice although a parameter labeling nonequivalent
central extensions has been taken to be equal $2k$. Even for
$\gcd(k,N)=1$ we have $\gcd(2k,2N)=2$, so the condition accompanying
\req{Bcon} that $l$ is mutually prime with the period, is not
fulfilled in this case. This leads to the magnetic periodicity with a
period $N$ though the lattice (crystal) period is $2N$. This problem
was briefly discussed by Brown~\cite{brown} and Zak~\cite{zak1b} and
its solution is possible through a concept of {\em magnetic cells}
--- if $\gcd(l,N)=\lambda>1$ [in \req{Bcon} and \req{Zcon}] then the
$N\times N$ lattice can be decomposed into
$(N/\lambda)\times(N/\lambda)$ magnetically periodic sublattices,
which form a $\lambda\times\lambda$ lattice of magnetic cells. In the
case considered $\lambda=2$ and the $(2N)\times(2N)$ lattice is
decomposed into four $N\times N$ sublattices consisting of points
$[X,Y]$, $[X+N,Y]$, $[X,Y+N]$, and $[X+N,Y+N]$, respectively, where
$X,Y\in{\Bbb Z}_N$. According to \req{Frep} and \req{Ffac2} an
irreducible projective representation of $T_{2N}$ should be
$N$-dimensional in this case. It follows from the Burnside's theorem
that there are {\em four}\/ such representations and they can be
chosen as
 \begin{equation}\label{Frep2}
 {^{\,l}_3}D^{\kappa_x,\kappa_y}_{jk}[X,Y]=
 (-1)^{\kappa_x\epsilon_x+\kappa_y\epsilon_y}
 \exp\biggl(2\pi\ii\frac{l}{N}Xj\biggr)\delta_{j,k-Y}\,,
 \end{equation}
 where $\kappa_x,\kappa_y=0,1$ and $\epsilon_x$ ($\epsilon_y$) is
equal to 0 for $X<N$ ($Y<N$) and to 1 otherwise. Therefore the
representations considered by Brown and Zak \req{BZrep} are
equivalent to ${^{\,l}_3}D^{0,0}$. However, the latter is clearly
periodic with the period $N$ and satisfies the condition
\req{strongx}. Since the trivial factor system \req{tfac2} is not
normalized in $C_N$ then the representations \req{BZrep} satisfy the
condition \req{strong} and do not satisfy \req{strongx}. 

\section{Discussion and final remarks}\label{Dis}

Summarizing the above discussion on odd and even periods $N$ we can
state that the antisymmetric gauge $\vA=[Hy,-Hx]/2$, considered by
Brown and Zak, corresponds in fact to the crystal period $2N$ and the
magnetic period $N$. If one, like Brown and Zak, does not take this
fact into account then results obtained can lead to erroneous
conclusions. For example, in this way the additional condition
\req{Zcon} was derived. Investigations of the standard and normalized
factor system \req{Ffac}, corresponding to the Landau gauge, have
clearly indicated points at which the magnetic translation groups are
`gauge-dependent' and how one can `normalize' factor systems and
representations.

The magnetically periodic boundary conditions of projective
representations, when the Landau gauge is assumed, can be invoked if
 \begin{equation}\label{Fcon}
 H={l\over N}\varphi_0\,,
 \end{equation} 
 where $l=0,1,\dots,N-1$. (If $\gcd(l,N)=\lambda$ then the magnetic
period is equal to $N/\lambda$, whereas the crystal period is still
$N$.) The factor system \req{Ffac} (and also the representation
\req{Frep} and the physical properties) is a periodic function of $H$
with a period $\varphi_0$. Hence the different magnetic response of
the considered system can be observed only for $N$ values of
$H=l\varphi_0/N$. 

If $N$ is an odd integer then $\gcd(2l,N)=\gcd(l,N)$ so the magnetic
periodicity is the same in both cases and $l$ in \req{Fcon} can be
replaced by $2l'$, which is equivalent to Zak's condition \req{Zcon}.
However, the successively counted values of $HN/\varphi_0$ have to be
arranged in a different order: $0,N+1,2,N+3,\dots,2N-2,N-1$. If these
values were arranged in the increasing order, \ie,
$0,2,\dots,N-1,N+1,\dots,2N-2$, it might suggest that the condition
\req{Zcon} has to be taken into account and that the magnetic period
is $2\varphi_0$. The only way to settle this problem is by
investigation of a system described by a Hamiltonian with a
nonperiodic part, \eg, the paramagnetic term. 
 
The case of even $N$ has a quite different nature. As was shown above,
the factor systems and representations considered by Brown and Zak
describe a lattice with the crystal period $2N$ and the magnetic
period equal $N$. The condition \req{Fcon} yields
$H=l\varphi_0/(2N)$, with $l=0,1,\dots,2N-1$, but to achieve the
magnetic period $N$ only even values of $l=2l'$ are considered, so
$H=l'\varphi_0/N$. The representation \req{Brep} [see also
\req{BZrep}] is one of four nonequivalent irreducible representations
which can be determined in this case. It can be easily seen that
Brown's considerations for odd $N$ can also be interpreted in this
way (since the decomposition of a $(2N)\times(2N)$ lattice into four
$N\times N$ lattices does not depend on the parity of $N$). It means,
in particular, that the antisymmetric gauge for $N=2$ can be
introduced only if one considers the $4\times 4$ lattice with $H=0$
(a trivial case) or $H=\varphi_0/2$. 

In this work the descriptions of the magnetic translation group
proposed by Brown and Zak were compared with the results obtained by
means of the Mac~Lane method.~\cite{florek94} The first authors
assumed the antisymmetric gauge, whereas the Mac~Lane method led to
the Landau gauge. Due to a factor ${1\over2}$ in the antisymmetric
gauge some problems arise when one introduces the magnetically
periodic boundary conditions. More careful considerations put forward
by Zak gave the additional condition \req{Zcon} for an odd period
$N$. However, the condition \req{Fcon} obtained for the Landau gauge
resembles the Brown's condition \req{Bcon} and does not depend on the
parity of $N$. This condition was obtained from the group-theoretical
considerations leading to the factor system \req{Ffac}. In the next
step $k/N$ was interpreted as $H/\varphi_0$. At first sight it can be
interpreted as any value proportional to $H$, \eg, as $2H/\varphi_0$.
However, the first choice is confirmed by the value of
group-theoretical commutator, which does not depend on the gauge or,
in the other words, is identical for all equivalent factor systems. 

Let us also remind that in this work $H$ in fact denotes the magnetic
flux through one primitive cell. Therefore, according to \req{Bcon}
or \req{Fcon}, the total flux through the $N\times N$ lattice is 
equal to
 \begin{equation}\label{Fi}
 \Phi=lN\varphi_0\,,
 \end{equation}
 \ie, to an integer multiplicity of the fluxon. To introduce the
antisymmetric gauge one has to consider a $(2N)\times(2N)$ lattice
and even $l=2l'$. Hence the total flux equals $\Phi=4l'N\varphi_0$,
so the flux through one $N\times N$ magnetic cell is equal to
$l'N\varphi_0$, which is consistent with teh previous value \req{Fi},
and the flux trough one primitive cell is equal to $H_a=l'\varphi_0/N$.
On the other hand, the flux through one primitive cell of the
$(2N)\times(2N)$ lattice (assuming the Landau gauge) is
$H_L=l\varphi_0/(2N)$, so in general it is a half of $H_a$.
Therefore we can set up the certain procedure: For a given magnetic
field $H_a$ the antisymmetric gauge can be introduced if the
magnetically periodic boundary conditions admit two times smaller
$H_L$. For the sake of illustration let us consider the 
$(2N)\times(2N)$
lattice and $H=\varphi_0/N$. Then, from \req{BZcom}, one obtains a
commutator corresponding to the primitive vectors $[1,0]$ and $[0,1]$
as 
 $$
 c([1,0],[0,1])=\exp(-2\pi\ii/N)\,.
 $$ 
 The formula \req{BZfac} gives the following values of the
corresponding factors [see \req{comm}]
 $$
 {_1}m([1,0],[0,1]) = \exp(-\pi\ii/N)
 $$
 and
 $$
 {_1}m([0,1],[1,0]) = \exp(\pi\ii/N)\,,
 $$
 whereas \req{Ffac} leads to
 $$
 {_2}m([1,0],[0,1]) = 1
 $$
 and
 $$
 {_2}m([0,1],[1,0]) = \exp(2\pi\ii/N)\,.
 $$
 So, the flux through the primitive cell, corresponding to the
commutator and independent of the gauge, was decomposed in two
different ways into fluxes through `primitive' triangles. However,
the first decomposition (related to the antisymmetric gauge) is not
possible for the minimal flux $H=\varphi_0/(2N)$. If considering any
other trivial factor system \req{phiw} determined by the parameter
$w\in{\Bbb R}$ one can obtain many other decompositions of the
commutator into factors. It can be viewed as the decomposition of the
flux through the primitive cell into fluxes through the `lower' and
`upper' triangle. In particular, the other Landau gauge,
corresponding to the factor system \req{Ffacbar}, changes roles of
these triangles since one obtains
 $$
 \overline{m}([1,0],[0,1]) = \exp(2\pi\ii/N)
 $$
 and
 $$
 \overline{m}([0,1],[1,0]) = 1\,.
 $$


Despite the fact that the physical properties are gauge-independent
we have noticed that the form of the vector potential \vA\ has a
certain importance in the mathematical description of a system. 
One has
to be especially very careful considering projective
representations or extensions of groups, since some equivalent factor
systems are neither standard nor normalized. However, it may occur
that in certain applications or in other descriptions of the same
problem it is more convenient to use such a form of \vA.


\begin{thebibliography}{99}
\bibitem{m3n3} Since a two-dimensional lattice is considered then a
term $m_3n_3/N$ is omitted in \req{Zrep} as compared with Zak's
formula and the finite magnetic translation group consisting of
$bN^3$ elements is considered [$b$ is defined in the text below
\req{Zrep}].
\end{thebibliography}
\end{document}